\documentclass[twocolumn]{aastex631}

\usepackage{amsmath}
\usepackage{bm}
\usepackage{verbatim}

\shortauthors{Deng et al.}

\begin{document}

\title{An Old, Low-mass, Metal-poor Hypervelocity Star Candidate Consistent with a Galactic Center Origin}

\correspondingauthor{Yang Huang}
\email{huangyang@ucas.ac.cn}

\author[0009-0008-5286-1060]{Shunhong Deng}
\affiliation{School of Astronomy and Space Science, University of Chinese Academy of Sciences, Beijing 100049, China}
\affiliation{School of Physics and Astronomy, China West Normal University, No. 1 Shida Road, Nanchong 637009, China}

\author[0000-0003-3250-2876]{Yang Huang}
\affiliation{School of Astronomy and Space Science, University of Chinese Academy of Sciences, Beijing 100049, China}
\affiliation{National Astronomical Observatories, Chinese Academy of Sciences, Beijing 100101, China}

\author[0009-0002-4542-8046]{Haozhu Fu}
\affiliation{Department of Astronomy, School of Physics, Peking University, Beijing, 100871, China}
\affiliation{Kavli Institute for Astronomy and Astrophysics, Peking University, Beijing, 100871, China}

\author[0000-0002-3935-2666]{Yongkang Sun}
\affiliation{National Astronomical Observatories, Chinese Academy of Sciences, Beijing 100101, China}
\affiliation{School of Astronomy and Space Science, University of Chinese Academy of Sciences, Beijing 100049, China}

\author{Qikang Feng}
\affiliation{Department of Astronomy, School of Physics, Peking University, Beijing, 100871, China}
\affiliation{Kavli Institute for Astronomy and Astrophysics, Peking University, Beijing, 100871, China}

\author[0009-0004-8011-4207]{Guoyang Chen}
\affiliation{Leiden Observatory, Leiden University, PO Box 9513, 2300 RA Leiden, The Netherlands}
\affiliation{National Astronomical Observatories, Chinese Academy of Sciences, Beijing 100101, China}

\author[0000-0002-7727-1699]{Huawei Zhang}
\affiliation{Department of Astronomy, School of Physics, Peking University, Beijing, 100871, China}
\affiliation{Kavli Institute for Astronomy and Astrophysics, Peking University, Beijing, 100871, China}


\begin{abstract}
We report the discovery of DESI-HVS1, a hypervelocity star (HVS) candidate identified from DESI DR1 spectroscopy and Gaia DR3 astrometry. DESI-HVS1 is an old, low-mass, metal-poor F-type star with a mass of $0.8\,M_\odot$, an age of $\sim14.1$~Gyr, and $\mathrm{[Fe/H]}=-1.6$. It is located at a heliocentric distance of $3.77^{+0.39}_{-0.36}$~kpc and has a Galactocentric total velocity of $523^{+46}_{-47}\,\mathrm{km\,s^{-1}}$, marginally exceeding the local escape speed, corresponding to an unbound probability of $P_{\rm ub} \sim 50\%$.
Backward orbit integrations show that DESI-HVS1 had a closest approach to the Galactic center (GC) of $0.40^{+0.23}_{-0.11}\,\mathrm{kpc}$, with a velocity of $682^{+22}_{-35}\,\mathrm{km\,s^{-1}}$ and a flight time of $12.89^{+0.92}_{-0.74}\,\mathrm{Myr}$.
The reconstructed orbit exhibits a clear perigalactic turning point and only a single crossing of the Galactic midplane ($P_{\rm cross} > 0.95$).
These properties suggest that DESI-HVS1 is most naturally explained by the Hills mechanism, although alternative scenarios cannot be entirely ruled out. Its discovery provides the first strong evidence for an old, low-mass HVS candidate consistent with a GC origin, indicating that the apparent dominance of young, massive GC-origin HVSs is likely a consequence of observational selection effects.
\end{abstract}
\keywords{Hypervelocity stars (776) --- Stellar dynamics (1596) --- Stellar kinematics (1608) --- Galactic center (565) --- Supermassive black holes (1663)}

\section{Introduction} 
\label{Sect.1}
\par The Galactic center (GC), hosting the central supermassive black hole \citep[SMBH;][]{Ghez2003}, is one of the most dynamically extreme environments in the Milky Way (MW). The SMBH's strong gravitational potential, together with the extremely high stellar density in the surrounding GC region, produces frequent dynamical interactions. Among these interactions, one particularly striking process is the three-body encounter between the SMBH and a stellar binary, in which one star can be captured while the other is ejected at velocities exceeding $1000\,\mathrm{km\,s^{-1}}$, sufficient to escape the Galaxy. Stars ejected in this way are known as hypervelocity stars (HVSs), a scenario first proposed by \citet{Hills1988} and commonly referred to as the Hills mechanism.
Subsequent work has explored additional GC-related channels for producing HVSs, including interactions involving a binary SMBH system \citep{Yu2003,Cao2025} and encounters between the SMBH and infalling globular clusters \citep{CapuzzoDolcetta2015}. Beyond the GC, HVSs may also be produced by supernova explosions in close binaries \citep[e.g.,][]{Blaauw1961,PortegiesZwart2000}, dynamical encounters in dense stellar systems \citep[e.g.,][]{Leonard1990,Gvaramadze2009}, tidal stripping during the accretion and disruption of satellite galaxies \citep[e.g.,][]{Abadi2009,Huang2021}, or Hills-like ejections in globular clusters hosting intermediate-mass black holes \citep[IMBHs;][]{Huang2025}.

\par Since the discovery of the first HVS by \citet{Brown2005}, numerous candidates have been reported \citep[e.g.,][]{Zheng2014,Brown2014,Huang2017,Sun2025}. However, few can be directly linked to a GC origin, as uncertainties in distances and proper motions limit the precision of backward orbit integrations for these distant HVSs.
A notable exception is the metal-rich A-type star S5–HVS1 from the Southern Stellar Stream Spectroscopic Survey (S$^5$), whose total velocity of $1755\pm50~\mathrm{km\,s^{-1}}$ and backward trajectory unambiguously point to the GC \citep{Koposov2020}, providing compelling evidence for the Hills mechanism. More recently, \citet{Liao2023} identified two additional HVS candidates of GC origin based on detailed kinematic analyses using Gaia Data Release 3 (DR3). Complementary efforts have also focused on high-velocity bound stars that may have been ejected from the GC but do not necessarily exceed the Galactic escape speed \citep[e.g.,][]{Cavieres2026}.
Strikingly, most candidate GC-origin HVSs are young, massive, early-type stars, including the only unambiguous case identified so far, S5–HVS1. No old, low-mass, late-type HVSs from the GC have yet been confirmed, leaving this population conspicuously underrepresented. This notable lack is surprising under a standard initial mass function (IMF), which would predict many more low-mass GC-ejected HVSs. Two explanations have been proposed: either the GC stellar population is intrinsically top-heavy, favoring massive stars, or the scarcity reflects observational selection effects, as faint, low-mass HVSs are challenging to detect in current surveys \citep{Brown2015}.
Understanding this issue ultimately hinges on whether and how many low-mass, late-type GC-origin HVSs can be found.

\par In this Letter, we report the discovery and detailed analysis of DESI-HVS1, the first candidate GC-origin HVS that is old, low-mass, and late-type, identified using spectroscopy from DESI DR1 \citep{DESI2025} and precise astrometry from Gaia DR3 \citep{gaia2023}. 
This Letter is organized as follows. In Section~\ref{Sect.2}, we present the observational data and derived properties of DESI-HVS1. Section~\ref{Sect.3} explores its possible origins through backward orbital and dynamical analyses. Finally, we summarize our findings in Section~\ref{Sect.4}.

\setlength{\tabcolsep}{0.5pt} 
\begin{deluxetable}{lcc}
\tablecaption{Detailed information of the HVS candidate DESI-HVS1.}
\label{tab1}
\tablehead{Parameter & Value  & Units}
\startdata
R.A. (J2000) &16:07:59.05& \\
Decl. (J2000)&$+$33:50:35.27 & \\
Gaia DR3 source\_id   & 1323614710521141120   &   \\
Gaia DR3 Proper motion $\mu_{\alpha}\cos\delta$&$-18.545 \pm 0.046$&mas\,yr$^{-1}$\\
Gaia DR3 Proper motion $\mu_{\delta}$&$20.904 \pm 0.056$&mas\,yr$^{-1}$\\
Gaia DR3 Parallax&$0.272 \pm 0.051$&mas\\
Gaia DR3 $G$-band magnitude&$16.800 \pm 0.001$&mag\\
Gaia DR3 $G_{\rm BP} - G_{\rm RP}$&$0.718 \pm 0.007$&mag\\
RUWE&1.01& \\
Parallax-based Distance &$3.84^{+0.96}_{-0.65}$&kpc\\
Improved Distance &$3.77^{+0.39}_{-0.36}$&kpc\\
SDSS  $g$-band magnitude&$17.122 \pm 0.004$&mag\\
SDSS  $r$-band magnitude&$16.794 \pm 0.004$&mag\\
SDSS  $i$-band magnitude&$16.681 \pm 0.005$&mag\\
SDSS  $z$-band magnitude&$16.643 \pm 0.009$&mag\\
$E(B - V)_{\rm SFD}$& $0.019$ &mag\\
DESI DR1 target\_id   & 39632956725136475  &    \\
S/N   & 51.5 &    \\
Line-of-sight velocity $v_{\rm los}$&$-145.44\pm 0.81$&km\,s$^{-1}$\\
Effective temperature $T_{\rm eff}$&$6198 \pm 19$&K\\
Surface gravity log\,$g$&$4.77 \pm 0.05$&dex\\
Metallicity \text{[Fe/H]}&$-1.64\pm 0.03$&dex\\
$\alpha$-element to iron ratio \text{[$\alpha$/Fe]}&$0.51 \pm 0.03$&dex\\
Galactocentric distance $R$&$7.63 \pm 0.03$&kpc\\
Total velocity $V_{\rm GSR}$&$523.41^{+46.17}_{-46.79}$&km\,s$^{-1}$\\
$P\mathrm{_{cross}}$ & 1.00 & \\
$r\mathrm{_{closest}}$ & $0.40^{+0.23}_{-0.11}$ & kpc \\
$v\mathrm{_{closest}}$ & $682.35^{+21.63}_{-34.66}$ & km\,s$^{-1}$ \\
$t\mathrm{_{fly}}$ & $12.89^{+0.92}_{-0.74}$ & Myr \\
Age&$14.1^{+1.0}_{-1.5}$&Gyr\\
Mass&$0.76^{+0.03}_{-0.02}$&$M_\odot$\\
\enddata
\tablecomments{The improved distance is derived from isochrone fitting using photometric and spectroscopic constraints, with the parallax distance adopted as an input prior (Section \ref{Sect.2.3}). $P_{\rm cross}$ is the probability that the star undergoes a single crossing of the Galactic midplane. 
$r_{\rm closest}$ is the minimum Galactocentric distance attained during backward orbit integration, 
with $v_{\rm closest}$ and $t_{\rm fly}$ giving the Galactocentric total velocity at that closest approach and the corresponding flight time. 
All quantities are computed using the {\tt galpy} {\tt MWPotential2014}.}
\end{deluxetable}

\section{Data and Stellar Properties}
\label{Sect.2} 

\subsection{Spectroscopy}
\label{Sect.2.1}
\par
\begin{figure}[!t]
\centering
\includegraphics[width=1.025\linewidth]{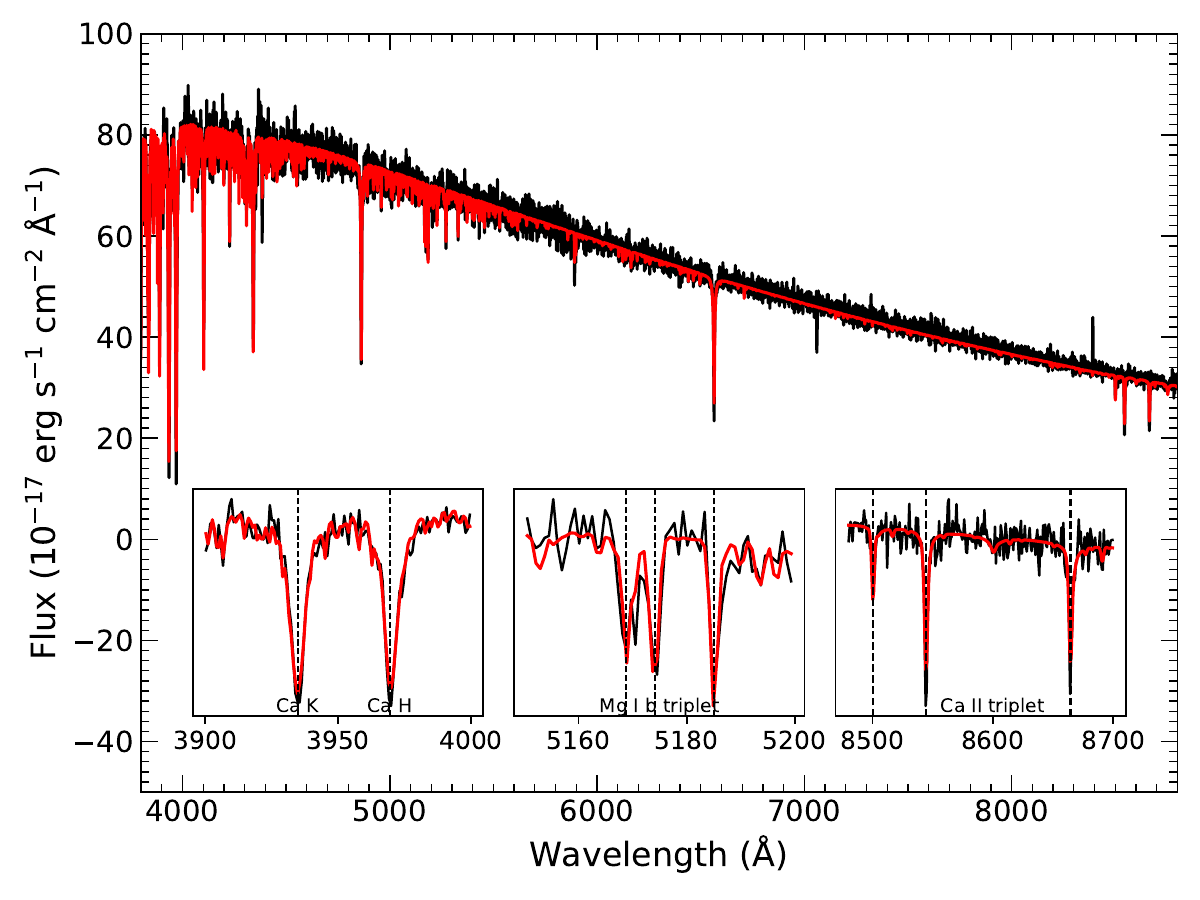}
\caption{The observed spectrum (black) is compared with a synthetic spectrum (red) from the Göttingen spectral library \citep{Husser2013}, computed with $T_{\rm eff}=6200\,\mathrm{K}$, $\log g=4.5$, $[\mathrm{Fe/H}]=-1.5$, and $[\alpha/\mathrm{Fe}]=0.40$. Insets show enlarged views of the Ca\,\textsc{ii} K $\lambda3933$ and H $\lambda3968$, the Mg\,\textsc{i}\,\textit{b} triplet $\lambda\lambda5167,5173,5184$, and the Ca\,\textsc{ii} triplet $\lambda\lambda8498,8542,8662$ lines.}
\label{fig1}
\end{figure}

 The DESI DR1 provides high-quality spectra from the first year of its five-year survey \citep{DESI2025}, covering 3600–9800 \AA\ at $R \sim 2000$–5000. Stellar parameters are derived with the DESI Stellar Parameter Pipeline (SPP) using RVSpecFit \citep{Koposov2019} and FERRE \citep{Allende2006}, yielding effective temperature ($T_{\rm eff}$), surface gravity ($\log g$), metallicity ([Fe/H]), alpha-element abundance ([$\alpha$/Fe]), and line-of-sight velocity ($v_{\rm los}$). 
DESI-HVS1 meets stringent quality criteria, including \texttt{SUCCESS = True}, \texttt{RVS\_WARN = 0}, \texttt{FEH\_ERR < 0.2}, and \texttt{RR\_SPECTYPE = STAR}. The \texttt{RVSpecFit} pipeline classifies DESI-HVS1 as an F-type star with $T_{\rm eff} = 6198 \pm 19$ K, $\log g = 4.77 \pm 0.05$ dex, [Fe/H] $= -1.64 \pm 0.03$ dex, and $v_{\rm los} = -145.44 \pm 0.81$ km s$^{-1}$ from a spectrum with Signal-to-Noise Ratio (S/N) = 51.5 (see Table~\ref{tab1}).
A comparison between the observed spectrum of DESI-HVS1 and a synthetic spectrum from the G\"ottingen Spectral Library \citep{Husser2013} with similar atmospheric parameters shows excellent agreement (Figure~\ref{fig1}), supporting the reliability of the derived stellar parameters.

\begin{figure*}[!ht]
\centering
\includegraphics[width=0.85\linewidth]{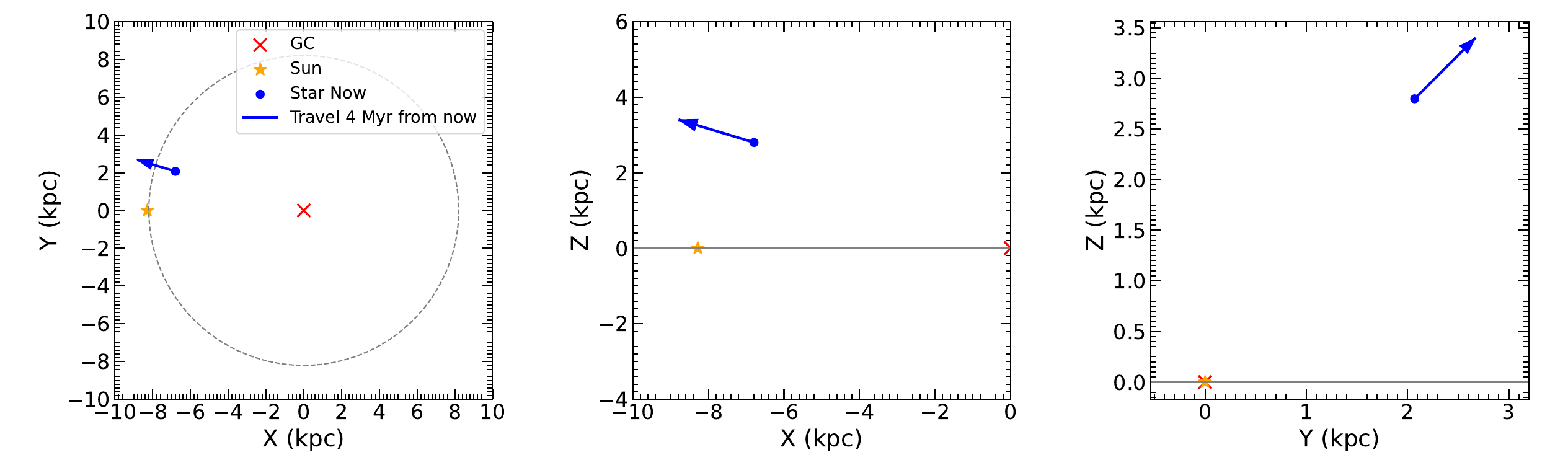}
\caption{Location and direction of motion of DESI-HVS1 in the Galaxy. The three panels show different projections in Galactic Cartesian coordinates. The positions of the Sun, the GC, and the solar circle are indicated by the orange star, red cross, and gray circle, respectively. The blue line indicates the path DESI-HVS1 will travel over the next 4 Myr. The arrows show the velocity directions in each projection.}
\label{fig2}
\end{figure*}

\begin{figure}[!t]
\centering
\includegraphics[width=1.025\linewidth]{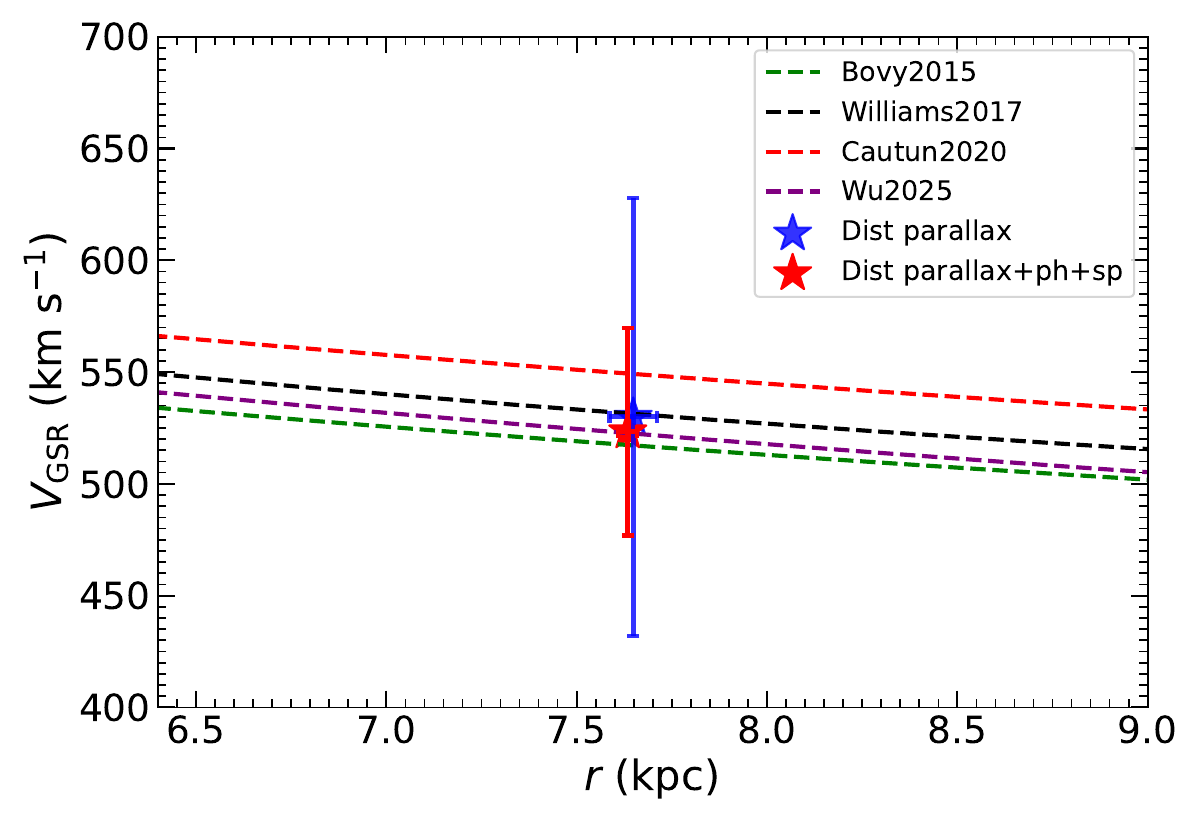}
\caption{Galactocentric total velocity of DESI-HVS1 compared with escape velocity curves. 
The black dashed line shows the observed Galactic escape velocity curve from \citet{Williams2017}. 
The green, red, and purple dashed lines correspond to predictions from the Galactic potential models (\citealp{Bovy2015}, \citealp{Cautun2020}, and \citealp{Wu2025}, respectively). 
The blue star shows the result obtained using the parallax-based distance ($R = 7.65 \pm 0.06~{\rm kpc}$, $V_{\rm GSR} = 530^{+100}_{-96}~{\rm km\,s^{-1}}$), while the red star corresponds to the improved distance ($R = 7.63 \pm 0.03~{\rm kpc}$, $V_{\rm GSR} = 523^{+46}_{-47}~{\rm km\,s^{-1}}$).}
\label{fig3}
\end{figure}

\subsection{Astrometry}
\label{Sect.2.2}
The Gaia DR3 \citep{gaia2023} provides high-quality astrometry for DESI-HVS1, with 
$\texttt{RUWE} < 1.4$, 
$\sigma_\varpi / \varpi < 0.2$, 
and $\varpi - \varpi_{\rm zp} > 0$ \citep{Lindegren2021b}. 
The heliocentric distance is inferred from the Gaia DR3 parallax after applying the zero-point correction $\varpi_{\rm zp}$, 
which depends on ecliptic latitude, magnitude, and stellar color and is computed following \citet{Lindegren2021b}. 
We adopt a Bayesian framework to infer the distance and its uncertainty by incorporating the parallax measurement and its error:
\begin{equation}
\label{eq1}
P(d|\varpi,\sigma_{\varpi}) \propto P(\varpi|d,\sigma_{\varpi})\, d^2\, P(r_{\rm GC}) \text{,}
\end{equation}
\begin{equation}
P(\varpi|d,\sigma_{\varpi}) = 
\frac{1}{\sqrt{2\pi}\sigma_{\varpi}} 
\exp\left[-\frac{(\varpi-\varpi_{\rm zp}-1/d)^2}{2\sigma_\varpi^2}\right] \text{,}
\end{equation}
where $r_{\rm GC}$ is the Galactocentric distance. 
For the spatial prior, we assume a halo-like power-law density profile with slope $\alpha = 3.39$ \citep{McMillan2017}. 
Using this method, we infer a heliocentric distance of $3.84^{+0.96}_{-0.65}$\,kpc for DESI-HVS1. This estimate is further refined by incorporating additional constraints from the Sloan Digital Sky Survey (SDSS) optical photometry and DESI spectroscopic chemical information, including metallicity and [$\alpha$/Fe], within a Bayesian framework, as described in the following Section~\ref{Sect.2.3}.
We note that the geometric distance from \citet{Bailer-Jones_2021}, $d = 3.45^{+0.93}_{-0.57}\,\mathrm{kpc}$, is consistent with our estimate within $1\sigma$, and our main conclusions remain unchanged when adopting this value.
\par We further derive the 3D position and velocity of DESI-HVS1 in the Galactocentric frame using the inferred heliocentric distance, the line-of-sight velocity $v_{\rm los}$ from DESI DR1, and the positions and proper motions from Gaia DR3. 
The Galactocentric total velocity is defined as
\begin{equation}
\label{eq3}
V_{\rm GSR} = \sqrt{V_X^2 + V_Y^2 + V_Z^2},
\end{equation}
where $(V_X, V_Y, V_Z)$ are the velocity components in the Galactocentric Cartesian coordinate system. 
Here, $X$ points from the Sun toward the GC, $Y$ is aligned with the direction of Galactic rotation, and $Z$ points toward the north Galactic pole.
We adopt a solar peculiar motion relative to the local standard of rest of $(U_\odot, V_\odot, W_\odot) = (11.69, 10.16, 7.67)\,{\rm km\,s^{-1}}$ \citep{Wang2021}, a circular velocity at the solar radius of $V_c(R_0) = 230\,{\rm km\,s^{-1}}$ \citep{Zhou2023}, a Galactocentric distance of the Sun $R_0 = 8.28\,{\rm kpc}$, and a vertical offset of the Sun from the Galactic midplane of $Z_\odot = 0.025\,{\rm kpc}$ \citep{BlandHawthorn2016, GravityCollaboration2024}. 
Uncertainties\footnote{For both distances, uncertainties are assumed to follow Gaussian distributions, with $\sigma$ approximated as $(d_{84\mathrm{th}}-d_{16\mathrm{th}})/2$.} are estimated via Monte Carlo (MC) simulations by randomly sampling  both the parallax-based distance and the improved heliocentric distance derived in Section~\ref{Sect.2.3}, together with the observational uncertainties in $v_{\rm los}$ (DESI DR1) and the proper motions (Gaia DR3). 
The resulting posterior distribution yields a Galactocentric total velocity of $V_{\rm GSR} = 529.96^{+99.61}_{-96.13}\ {\rm km\,s^{-1}}$ (using parallax-based distance) and 
$V_{\rm GSR} = 523.41^{+46.17}_{-46.79}\ {\rm km\,s^{-1}}$ (using improved distance), 
where the quoted uncertainties correspond to the 16th and 84th percentiles.
Using the improved distance, Figure~\ref{fig2} shows the present-day geometric configuration of DESI-HVS1 in Galactic Cartesian coordinates. The star is located at $(X,Y,Z)=(-6.79,\,2.08,\,2.81)\,\mathrm{kpc}$ and is moving away from the Galactic disk while continuing outward from the inner Galaxy.

\par As shown in Figure~\ref{fig3}, DESI-HVS1 lies at a Galactocentric radius of $R = 7.6$\,kpc. 
The total velocities inferred from both the parallax-based distance and the improved distance marginally exceed the escape speed ($V_{\rm esc}$) at this radius, as predicted by the Galactic escape velocity curve of \citet{Williams2017} and by the Galactic potential models of \citet[][i.e.~\texttt{galpy MWPotential2014}]{Bovy2015}, \citet{Cautun2020}, and \citet{Wu2025}. This suggests that DESI-HVS1 is an HVS candidate. For the adopted Galactic potential model, \texttt{galpy MWPotential2014}, the probability that the star is unbound is above 50\%. The corresponding probabilities for the other models or observational case are slightly lower but remain around 30--40\%.


\begin{figure}[!t]
\centering
\includegraphics[width=0.95\linewidth]{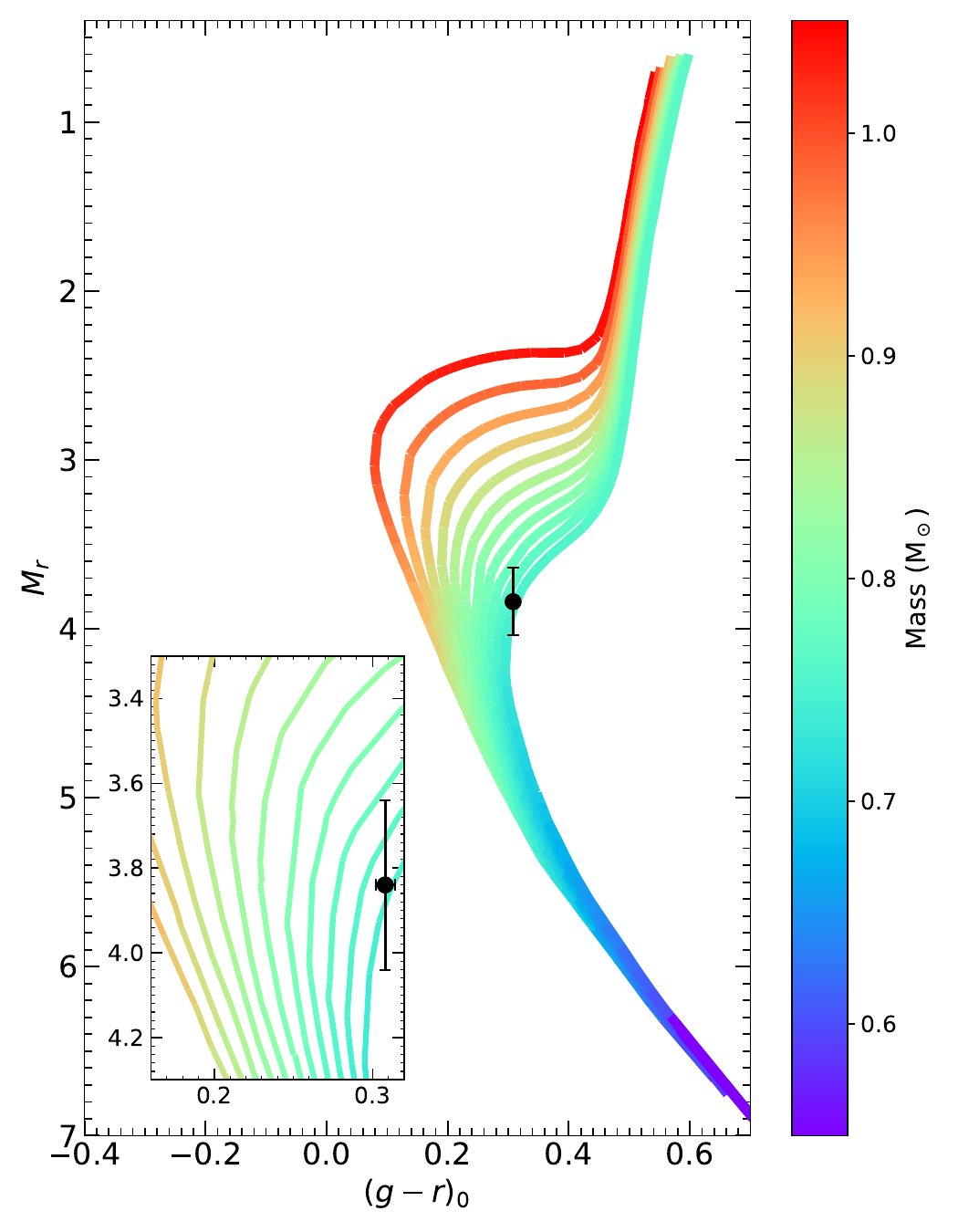}
\caption{Position of DESI-HVS1 on the $(g - i)_0$ versus $M_r$ diagram compared with theoretical stellar isochrones. The black point with error bars indicates the measured photometric location of DESI-HVS1. The background curves show PARSEC isochrones spanning ages from 5 to 15 Gyr in 1 Gyr increments, arranged from left to right in order of increasing age. The color scale denotes stellar mass, as illustrated by the color bar on the right. All isochrones are computed at fixed [M/H] = $-1.26$, converted from [Fe/H] = $-1.64$ and [$\alpha$/Fe] = $+0.50$. The inset panel provides a magnified view for clarity.}
\label{fig4}
\end{figure}

\subsection{Age and Mass}
\label{Sect.2.3}
The age and mass of DESI-HVS1 are derived using a Bayesian approach based on the observational constraints and stellar isochrones from the PARSEC stellar evolutionary models \citep{Bressan2012}. 
The observational constraints include the spectroscopic metallicity from DESI DR1, the SDSS photometric colors, and the absolute magnitudes. 
The SDSS magnitudes are corrected for interstellar extinction using the dust reddening from \citet{Schlegel1998}. 
The absolute magnitudes are inferred from the parallax-based distance derived above and the extinction-corrected SDSS photometry. 
In the estimation, the posterior probability distribution function (PDF) of the parameters is assumed to take the form:
\begin{equation}
\label{eq4}
f(\tau, m, M_\lambda) = N\,P(\tau, m)\,L(\tau, m),
\end{equation}
where  $\lambda =g, r, i, z$ and $N$ is the normalization constant that ensures the integral of 
$f(\tau, m, M_\lambda)$ over the parameter space equals unity.
The SDSS $u$ band is not used due to its large photometric uncertainty.
We adopt a uniform prior in age and assume a Salpeter IMF \citep{Salpeter1955} for the stellar mass. 
The likelihood function is defined as
\begin{equation}
L = \prod_{i=1}^{n} \frac{1}{\sqrt{2\pi}\sigma_i}
\exp\left(-\frac{\chi^2}{2}\right),
\end{equation}
where
\begin{equation}
\chi^2 = \sum_{i=1}^n 
\left( \frac{O_i - M_i(\tau, m)}{\sigma_i} \right)^2.
\end{equation}
Here, $O_i$ denotes the observed constraints (i.e., the intrinsic color $(g - r)_0$  and four SDSS bands' absolute magnitudes), and $M_i(\tau, m)$ represents the corresponding model predictions from the isochrones. 
We adopt ${\rm [M/H]} = -1.26$, which is converted from ${\rm [Fe/H]} = -1.64$ and ${[\alpha/{\rm Fe}]} = 0.50$ following \citet{Salaris1993}.

\par The age and mass are derived using the above approach with 5000 MC simulations. We adopt the 50th percentile of the posterior PDF as the best estimate, and the 16th and 84th percentiles as the corresponding uncertainties, yielding $\tau = 14.1^{+1.0}_{-1.5}\ \mathrm{Gyr}$, $M = 0.76^{+0.03}_{-0.02}\ M_\odot$.
From the four SDSS bands, we derive a refined weighted mean distance modulus of $\mu = 12.88^{+0.22}_{-0.22}$, corresponding to a refined heliocentric distance of $d = 3.77^{+0.39}_{-0.36}$\,kpc.
To assess the consistency of these results, we compare the location of DESI-HVS1 with theoretical stellar isochrones in the $M_r$ versus $(g-r)_0$ plane (Figure~\ref{fig4}), showing that the inferred parameters are well supported by the observational constraints. 

\section{Possible Origins of DESI-HVS1}
\label{Sect.3} 
\begin{figure}[!t]
\centering
\includegraphics[width=1.0\linewidth]{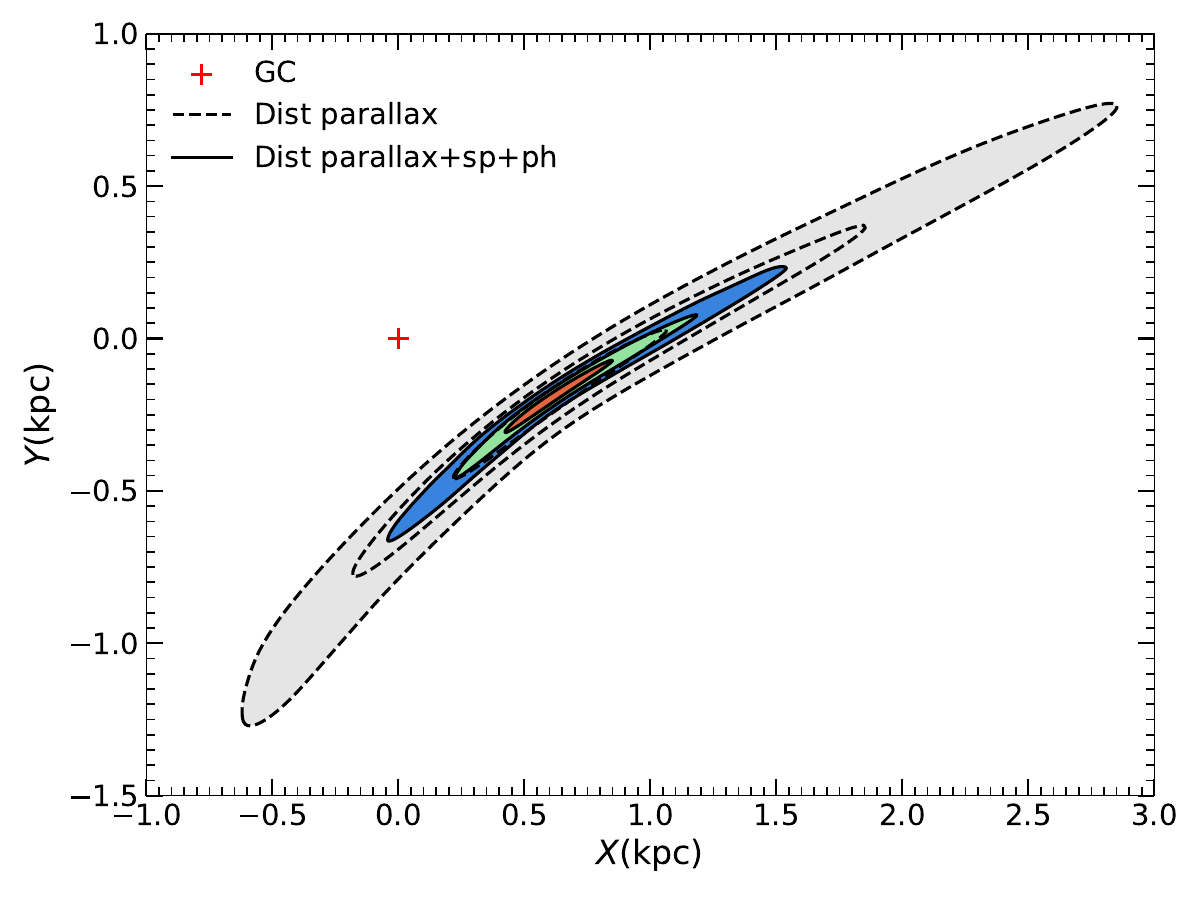}
\caption{The contours show the intersection regions of the backward-integrated orbits with the Galactic disk midplane in the $X$--$Y$-plane for the first crossing. The gray region is derived using the parallax-based distance, while the colored region is derived using the improved distance from isochrone fitting constrained by photometric and spectroscopic measurements. The dashed lines for the parallax-based distance and black solid lines for the improved distance from isochrone fitting mark the 16th, 50th, and 84th percentiles (from inner to outer) of the plane-crossing distribution, corresponding to the $1\sigma$ interval around the median.}
\label{fig5}
\end{figure}

\begin{figure*}[!t]
\centering
\includegraphics[width=0.85\linewidth]{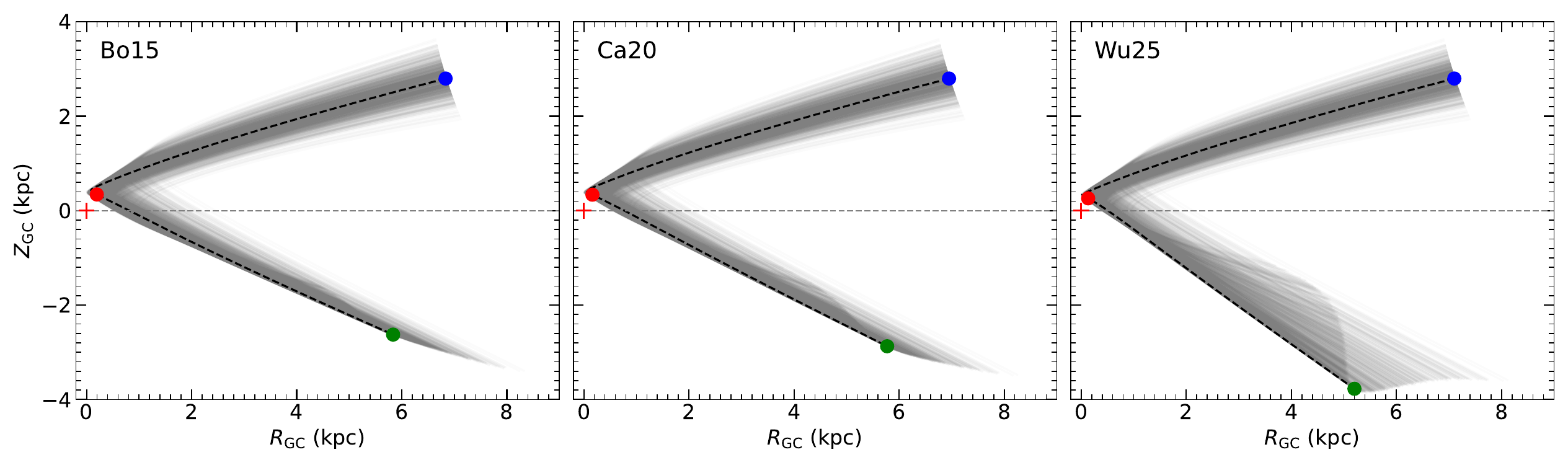}
\caption{The 1000 backward-integrated orbits of DESI-HVS1 in Galactocentric cylindrical coordinates ($R_{\rm GC}$, $Z_{\rm GC}$), computed under three different Galactic potential models and sampled from the uncertainties in its position and velocity. The black dashed curve shows the fiducial orbit. The blue, red, and green points represent the current position, the perigalacticon, and the position after ~50 Myr of backward orbital integration, respectively. The horizontal dashed line indicates the disk plane ($Z_{\rm GC}=0$\,kpc) and the GC at ($R_{\rm GC}, Z_{\rm GC}$) = (0, 0)\,kpc is marked by a red cross. The orbits are ballistic and nonclosed and exhibit a pronounced turning near perigalacticon.}
\label{fig6}
\end{figure*}

%
\subsection{GC Origin: Backward Orbit Analysis}
\label{Sect.3.1}
\par Several mechanisms have been proposed to eject HVSs with velocities significantly exceeding the typical random motions of halo stars ($\sim100$--$150~\mathrm{km\,s^{-1}}$; e.g., \citealt{Huang2016}). 
Among these, the Hills mechanism \citep{Hills1988} is widely invoked to explain HVSs ejected from the GC. 
To investigate whether DESI-HVS1 could have originated from the GC via the Hills mechanism, we integrate its orbit backward in time for $1~\mathrm{Gyr}$ with a time step of $0.1~\mathrm{Myr}$ using the {\tt galpy} package \citep{Bovy2015}. 
We adopt the {\tt MWPotential2014}, which consists of a power-law bulge potential, a Miyamoto--Nagai stellar disk \citep{Miyamoto1975}, and a Navarro--Frenk--White (NFW) dark matter halo \citep{NFW1996}. 
To test the robustness of our results, we also consider two alternative Galactic potential models, Ca20 and Wu25 \citep{Cautun2020,Wu2025}.

\par Figure~\ref{fig5} shows the intersection regions of the backward-integrated orbits of DESI-HVS1 with the Galactic disk midplane ($Z=0$\,kpc) in the $X$--$Y$-plane for the first crossing. 
Using the Gaia parallax distance from Equation~\ref{eq1}, the reconstructed orbit passes close to the GC, although the confidence contours are elongated due to the limited precision of the parallax measurement. Adopting the distance further refined by isochrone fitting (Equation~\ref{eq4}) significantly reduces the orbital uncertainty, yielding tighter confidence contours for the disk crossings. 
Using this improved distance for the orbit integration, DESI-HVS1 reaches its closest approach to the GC at $t_{\rm closest}=12.89^{+0.92}_{-0.74}~\mathrm{Myr}$ ago, with a minimum distance of $r_{\rm closest}=0.40^{+0.23}_{-0.11}~\mathrm{kpc}$ and a relative ejection velocity of $v_{\rm closest}=682.35^{+21.63}_{-34.66}\,\mathrm{km~s^{-1}}$. The corresponding three-dimensional position and velocity at that moment are 
$(X,Y,Z)=\left(0.14^{+0.17}_{-0.16},-0.03^{+0.31}_{-0.32},0.27^{+0.15}_{-0.13}\right)\,\mathrm{kpc}$ 
and 
$(V_X,V_Y,V_Z)=\left(-584^{+16.3}_{-16.6}, 178^{+15.1}_{-29.2}, 305^{+8.0}_{-30.8}\right)\,\mathrm{km~s^{-1}}$, respectively. 
The quoted uncertainties correspond to the 16th and 84th percentiles of the posterior distributions derived from 5000 MC realizations that incorporate uncertainties in the heliocentric distance, $v_{\rm los}$, and proper motions (following a similar procedure to estimate the uncertainty of $V_{\rm GSR}$ in Equation~\ref{eq3}). Consistent results are obtained when adopting two alternative Galactic potential models (Ca20 and Wu25; Table~\ref{tab2}), indicating that the inferred GC passage is not strongly sensitive to the assumed Galactic potential.

\par The fact that the orbit does not trace exactly back to the GC does not preclude a GC origin. The above analysis remains somewhat sensitive to both the adopted Galactic constants and the assumed dark matter halo shape. Adopting alternative sets of constants \citep[e.g.,][]{Reid2014,Huang2015} would shift the reconstructed plane-crossing contours in the $X$--$Y$-plane. In principle, one could adjust the Galactic constants, such as the solar position and velocity, to explore whether the resulting plane-crossing contours enclose the GC, similar to the approach applied in the analysis of S5-HVS1 \citep{Koposov2020}. 
Meanwhile, the Galactic potential adopted here does not include the effects of a nonspherical dark matter halo or perturbations induced by the Large Magellanic Cloud \citep{Kenyon2018,Boubert2020}. For example, the backward trajectories would be distorted if a triaxial dark matter halo were considered. In future work, we will combine DESI-HVS1 with other GC-origin HVSs and high-velocity stars to place joint constraints on both the Galactic constants and the shape of the MW dark matter halo.

\par Beyond the backward orbital reconstruction, DESI-HVS1 also satisfies key dynamical expectations for a GC-ejected HVS. Owing to its extreme ejection velocity, such a star is expected to follow a nearly ballistic trajectory and cross the Galactic midplane ($Z=0$\,kpc) only once over its orbital history. MC orbit integrations indicate $P_{\rm cross}>0.95$ under all three Galactic potential models, consistent with the criterion of \citet{Liao2023}.  Consistently, 1000 MC orbit integrations sampling the observational uncertainties under the three Galactic potential models recover trajectories that wrap around the GC, exhibit clear perigalactic turning points, and maintain a ballistic trajectory throughout the orbit (Figure~\ref{fig6}). Together, these dynamical tests further support a GC origin for DESI-HVS1.


\par The discovery of DESI-HVS1 as an old, low-mass HVS candidate consistent with a GC origin helps alleviate the long-standing puzzle that previously confirmed GC-ejected HVSs are almost exclusively massive and young. This result demonstrates that low-mass stars can also be ejected from the GC through the Hills mechanism, suggesting that the apparent dominance of high-mass HVSs may largely reflect selection effects rather than an intrinsically top-heavy IMF in the GC environment \citep{Marchetti2022}. 
Comparing DESI-HVS1 ($\sim0.8\,M_\odot$) with the well-established GC-origin HVS S5-HVS1 ($\sim2.35\,M_\odot$; \citealt{Koposov2020}), the current sample (one low-mass versus one high-mass HVS) implies $\alpha \approx 1.36$ under a simple power-law IMF scaling. Including DESI-312 ($\sim1.05\,M_\odot$; \citealt{Cavieres2026}), a bound high-velocity low-mass star potentially associated with the GC, increases the low- to high-mass count to two versus one, yielding $\alpha \approx 1.79$. These estimates remain highly tentative given the extremely limited sample size and the absence of corrections for selection biases or mass-dependent ejection efficiencies and should therefore be regarded as qualitative indications rather than quantitative constraints on the GC IMF.

\setlength{\tabcolsep}{5pt} 
\begin{deluxetable}{lcccc}
\tablecaption{Closest-approach properties of DESI-HVS1 under three different Galactic potential models.}
\label{tab2}
\tablehead{Potentials & $P\mathrm{_{cross}}$  & $r\mathrm{_{closest}}$&$v\mathrm{_{closest}}$ & $t\mathrm{_{fly}}$ \\
...& ... & (kpc) & (km\,s$^{-1}$)&(Myr)}
\startdata
Bo15 & 1.00 &    
$0.40^{+0.23}_{-0.11}$ &
$682.35^{+21.63}_{-34.66}$ &
$12.89^{+0.92}_{-0.74}$ \\
Ca20 & 0.98 & 
$0.43^{+0.24}_{-0.09}$ &
$705.63^{+16.15}_{-28.74}$ &
$12.53^{+0.84}_{-0.64}$ \\
Wu25 & 0.96 &
$0.36^{+0.24}_{-0.09}$ &
$771.84^{+12.97}_{-26.88}$ &
$12.64^{+0.91}_{-0.65}$ \\
\enddata
\end{deluxetable}

\subsection{Alternative Origins}
\label{Sect.3.2} 

\par In addition to the Hills mechanism, other dynamical interactions in the GC may also eject HVSs. For example, encounters between single stars and a binary SMBH or an SMBH–IMBH system can eject HVSs \citep{Yu2003}. However, the existence of such black hole binaries in the GC remains uncertain.

\par Ejection channels outside the GC also cannot be completely excluded. One possibility is acceleration through a supernova explosion in a binary system. Core-collapse supernovae typically impart velocities below $\sim300$--$400~{\rm km\,s^{-1}}$ to their companions \citep{PortegiesZwart2000}, which is insufficient to explain the velocity of DESI-HVS1. Thermonuclear supernovae may in principle produce HVSs with velocities approaching or exceeding $1000~{\rm km\,s^{-1}}$ \citep[e.g.,][]{Shen2018,Han2008,Bauer2019}. However, these scenarios require progenitor systems such as white dwarf (WD)+WD or WD+He-star binaries, and therefore the ejected companions are typically compact objects or hot subdwarfs. This is inconsistent with the stellar properties of DESI-HVS1. The Hills mechanism operating in globular clusters hosting IMBHs may also produce HVSs \citep{Fragione2019,Huang2025}. However, our backward orbit integrations do not show any association with the trajectories of known globular clusters, making a cluster origin unlikely.
\par Finally, we consider whether this star could have originated from the Gaia–Sausage–Enceladus (GSE; \citealt{Helmi2018}) merger. Although DESI-HVS1 resembles GSE stars in metallicity and follows a predominantly radial orbit, its orbital energy far exceeds that of typical GSE debris. In particular, its Galactocentric speed near or above the local escape speed, its robust backward trajectory passing within $\sim$ 0.4 kpc of the GC, and its single disk-plane crossing together disfavor a normal GSE origin.

\section{Summary}
\label{Sect.4}
\par In this Letter, we report the discovery of DESI-HVS1, an HVS candidate identified from DESI DR1 spectroscopy and Gaia DR3 astrometry. DESI-HVS1 is an old, low-mass, metal-poor F-type star located at a heliocentric distance of $3.77^{+0.39}_{-0.36}$~kpc, with a Galactocentric total velocity of $V_{\rm GSR}=523^{+46}_{-47}\,{\rm km\,s^{-1}}$, comparable to the escape speed at its position. Its present-day position and velocity vector indicate motion away from the Galactic disk and outward from the inner Galaxy. Backward orbit integrations using the {\tt galpy} potential {\tt MWPotential2014} \citep{Bovy2015} show that the star passed within $0.40^{+0.23}_{-0.11}$~kpc of the GC about $12.89^{+0.92}_{-0.74}$~Myr ago, with an inferred ejection velocity of $682^{+22}_{-35}\,{\rm km\,s^{-1}}$. MC orbit integrations sampling the observational uncertainties further show that the orbit remains strongly ballistic, exhibiting a clear turning point near perigalacticon and only a single crossing of the Galactic midplane ($Z=0$\,kpc) with a probability $P_{\rm cross} > 0.95$. These dynamical properties are naturally explained by the Hills mechanism operating in the GC, although alternative ejection scenarios cannot be completely excluded.

\par DESI-HVS1 therefore represents the first old, low-mass, metal-poor HVS candidate consistent with a GC origin through the Hills mechanism, extending the known population of GC-ejected HVSs beyond the previously identified young and massive stars. Its discovery suggests that the apparent scarcity of such objects likely reflects observational selection effects rather than an extremely top-heavy IMF in the GC environment. Continued advances in large-scale spectroscopic and astrometric surveys, particularly the forthcoming Gaia DR4, are expected to expand and refine the sample of GC-ejected HVS candidates, enabling deeper insights into the stellar demographics, physical conditions, and dynamical processes operating near the GC.
\section*{Acknowledgments}
\par We thank the anonymous referee for constructive comments. Y.H. acknowledges the support from the National Science Foundation of China (NSFC grant No. 12422303), the Fundamental Research Funds for the Central Universities (grant Nos. 118900M122, E5EQ3301X2, and E4EQ3301X2), and the National Key R\&D Programme of China (grant No. 2019YFA0405503). The DESI (\url{https://data.desi.lbl.gov/doc}) is managed by the U.S. Department of Energy, Office of Science, and carried out by an international collaboration of institutions. The DESI team acknowledges the sacred nature of Iolkam Du’ag (Kitt Peak) to the Tohono O’odham Nation and expresses gratitude for the opportunity to conduct observations on this ancestral land. This work has also made use of data from the European Space Agency (ESA) mission Gaia (\url{https://www.cosmos.esa.int/gaia}), processed by the Gaia Data Processing and Analysis Consortium (DPAC, \url{https://www.cosmos.esa.int/web/gaia/dpac/consortium}). Funding for the DPAC has been provided by national institutions, in particular those participating in the Gaia Multilateral Agreement.

\clearpage\newpage




\clearpage\newpage

\bibliography{deng25b}
\bibliographystyle{aasjournal}

\end{document}